\documentclass[a4paper]{article}

\usepackage{INTERSPEECH2015}

\usepackage{graphicx}
\usepackage{amssymb,amsmath,bm}
\usepackage{textcomp}
\usepackage{diagbox}

\ninept

\title{The SYSU System for the Interspeech 2015 Automatic Speaker Verification Spoofing and Countermeasures Challenge}

\makeatletter
\def\name#1{\gdef\@name{#1\\}}
\makeatother \name{{\em Shitao Weng$^1$, Shushan Chen$^1$, Lei Yu$^1$, Xuewei Wu$^1$, Weicheng Cai$^2$, Zhi Liu$^1$, Ming Li$^{1,2}$}}

\address{$^1$SYSU-CMU Joint Institute of Engineering, Sun Yat-Sen University, Guangzhou, China\\
    $^2$SYSU-CMU Shunde International Joint Research Institute, Guangdong, China\\
    {\small \tt liming46@mail.sysu.edu.cn}}

\begin{document}

  \maketitle
  \begin{abstract}
Many existing speaker verification systems are reported to be vulnerable against different spoofing attacks, for  example speaker-adapted speech synthesis, voice conversion, play back, etc.
In order to detect these spoofed speech signals as a countermeasure, we propose a score level fusion approach with several different i-vector subsystems. We show that the acoustic level Mel-frequency cepstral coefficients (MFCC) features, the phase level modified group delay cepstral coefficients (MGDCC) and the phonetic level phoneme  posterior probability (PPP) tandem features are effective for the countermeasure. Furthermore, feature level fusion  of these features before i-vector modeling also enhance the performance. A polynomial kernel support vector machine is adopted as the supervised classifier. In order to enhance the generalizability of the countermeasure, we also adopted the cosine similarity and PLDA scoring as one-class classifications methods.
By combining the proposed i-vector subsystems with the OpenSMILE baseline which covers the acoustic and prosodic information further improves the final performance. The proposed fusion system achieves 0.29\% and 3.26\% EER on the development and test set of the database provided by the INTERSPEECH 2015 automatic speaker verification spoofing and countermeasures challenge.
  \end{abstract}

\section{Introduction} \label{sec:intro}
\begin{figure*}[!ht]
    \centering
    \includegraphics[width=16cm]{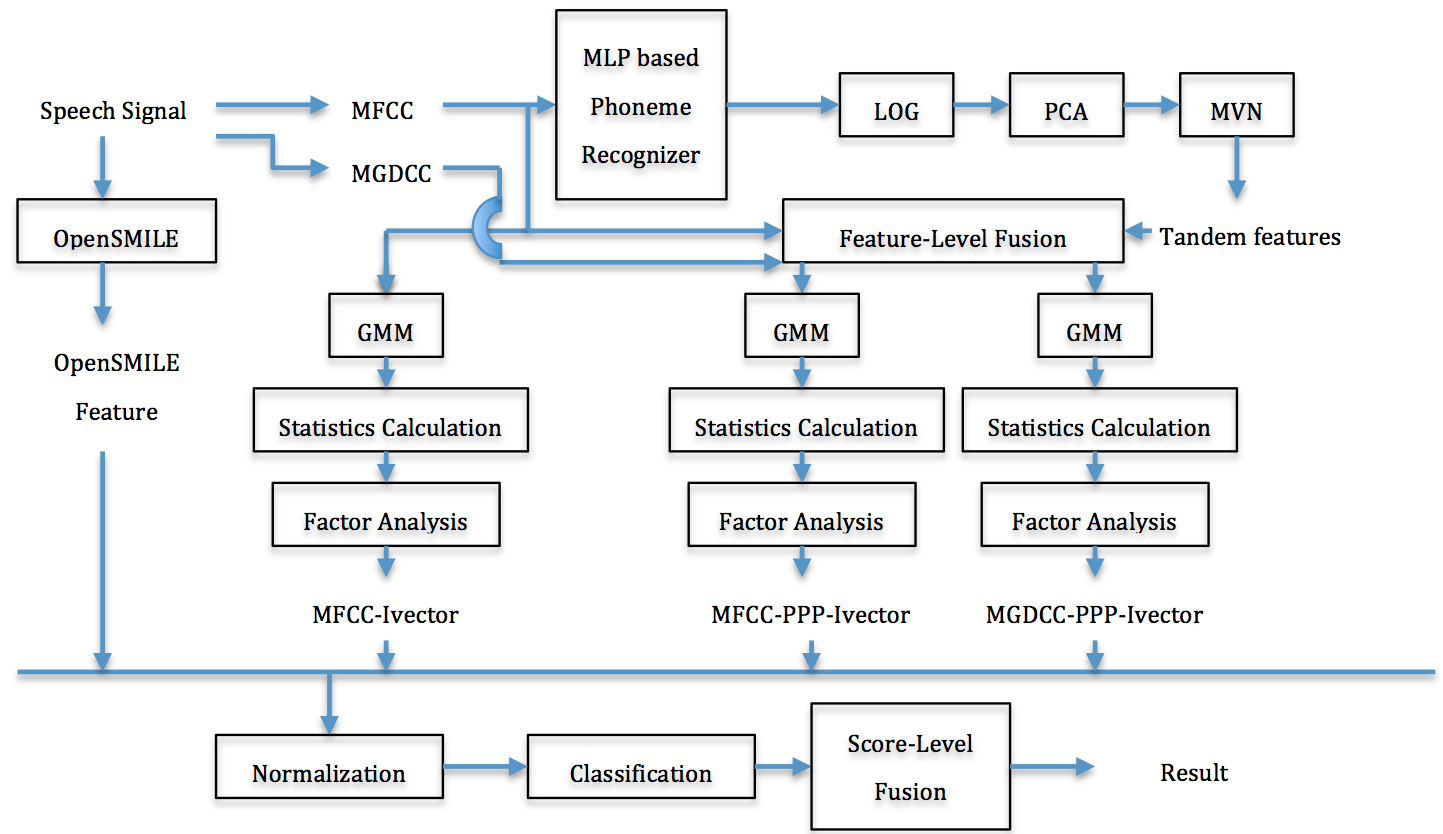}
    \caption{The system overview}\label{system-overview}
\end{figure*}
The goal of speaker verification  is to automatically verify the claimed speaker identity given a segment of speech. In the past decade, speaker verification has attracted significant research attention with promising results \cite{speakeridreview}. However, recently it is reported that many existing speaker verification systems are vulnerable against different spoofing attacks, e.g. speaker-adapted speech synthesis, voice conversion, play back, etc.\cite{DBLP:journals/speech/WuEKYAL15, DBLP:journals/taslp/YamagishiKNOI09, DBLP:conf/interspeech/WuVKCL13, DBLP:journals/taslp/TodaBT07,jointspeakeridandspoofing} 

Compared to text independent speaker verification, text dependent  speaker verification is more robust against  the play back spoofing   since  the speech content is constrained or  pre-defined. Speaker-adapted speech synthesis and voice conversion are the most common spoofing methods that can convert arbitrary text or speech inputs towards the target speaker \cite{DBLP:journals/speech/WuEKYAL15}.
To enhance the robustness of speech verification system against spoofing attacks, different countermeasures have been proposed. In \cite{alegre:hal-00783789}, higher-level dynamic features and voice quality assessment are used to detect those artificial signals. Furthermore, modified group delay cepstral coefficients (MGDCC) feature has been proposed to distinguish between the original and the spoofed speech signals in the phase domain \cite{wu2012detecting}. This approach is based on the fact that  the phase information of synthetic spoofing speech is typically different from the real human articulated speech while  the human auditory system is less sensitive to this difference. Long term temporal modulation feature derived from magnitude or phase spectrum has also been proposed to detect the synthetic speech \cite{DBLP:conf/icassp/WuXCL13}.

Total variability i-vector modeling has been widely used in speaker verification due to its excellent performance, compact representation and small model size \cite{dehak2011language,dehak2010front}. In this work, we apply the recently proposed generalized i-vector framework \cite{yun_icassp14,kennydeep,li2014interspeech,DHaro2014} with both the acoustic and phonetic features to the countermeasure task.

Figure \ref{system-overview} shows an overview of our anti-spoofing countermeasure system. First, there are several i-vector subsystems using different features, namely the acoustic level Mel-frequency cepstral coefficients (MFCC) features, the phase level MGDCC features, the phonetic level phoneme  posterior probability (PPP) tandem features \cite{li2014interspeech,li2014interspeechmbc} and their feature level combinations. Second, we also applied the openSMILE toolkit \cite{eyben2010opensmile} to perform the utterance level acoustic and prosodic feature extraction. We believe that the spoofed speech signal may have different prosodic patterns. Third, after the feature normalization, multiple classification methods, e.g. cosine scoring, K-nearest neighbor (KNN), simplified PLDA \cite{DBLP:conf/interspeech/Garcia-RomeroE11} and Support Vector Machine (SVM), are employed as the back end. Finally, score level fusion is performed to further enhance the overall system performance.

The remainder of the paper is organized as follows. The corpus and the proposed algorithms are explained in Sections \ref{sec-corpus} and \ref{sec-methods}, respectively. Experimental results and discussions are presented in Section \ref{sec-ex} while conclusions are provided in Section \ref{sec-con}.

\section{Corpus}\label{sec-corpus}

The database used to evaluate the proposed methods is based upon a standard dataset of both genuine and spoofed speech. Genuine speech is without significant channel or background noise effect and includes 106 speakers (45 male, 61 female), while spoofed speech is obtained through applying several spoofing algorithms on the genuine speech \cite{wu2014asvspoof}. The training data set (25 speakers, 3750 genuine utterances and 12635 spoofed utterances) is for model training while the development data set (35 speakers, 3497 genuine utterances and  49875 spoofed utterances) is used to evaluate the system performance and  tune the parameters. Finally, the testing data set (46 speakers, 193404 utterances) with unknown types of spoofing attacks is provided to obtain the official submission scores. The details of the  database and evaluation protocol are provided in \cite{wu2014asvspoof}.

\section{Methods}\label{sec-methods}
From Figure \ref{system-overview}, we can see that there are four different features, namely  MFCC i-vectors, MFCC-PPP i-vectors, MGDCC-PPP i-vectors and openSMILE feature vectors followed by the same feature normalization, classification  and score level fusion pipeline. We first present the proposed features in section \ref{sec:feature}. Then section \ref{sec:modeling} describes the supervised classification and score level fusion methods, respectively.

\subsection{Features}\label{sec:feature}

\subsubsection{The i-vector framework}\label{sec:ivector}

In the total variability space, there is no distinction between the speaker effects and the channel effects. Rather than  separately using the eigenvoice matrix $\mathbf{V}$ and the eigenchannel matrix $\mathbf{U}$ \cite{kenny2007joint}, the total variability space simultaneously captures the speaker and channel variabilities \cite{dehak2010front}.
Given a $C$ component GMM UBM model $\lambda$ with $\lambda_{c}=\{p_{c},\mathbf{\mu_{c}}, \mathbf{\Sigma_{c}}\},c=1,\cdots,C$ and an utterance with a $L$ frame feature sequence $\{\mathbf{y_{1}},\cdots,\mathbf{y_{L}}\}$, the zero-order and centered first-order Baum-Welch statistics on the UBM are calculated as follows:
\begin{equation}
N_{c}=\sum_{t=1}^{L}P(c|\mathbf{y_{t}},\lambda)
\label{eq_1}
\end{equation}
\begin{equation}
\mathbf{F_{c}}=\sum_{t=1}^{L}P(c|\mathbf{y_{t}},\lambda)(\mathbf{y_{t}}-\mathbf{\mu_{c}})
\label{eq_2}
\end{equation}
where $c=1,\cdots,C$ is the GMM component index and $P(c|\mathbf{y_{t}},\lambda)$ is the occupancy posterior probability for $\mathbf{y_{t}}$ on $\lambda_{c}$.
The corresponding centered mean supervector $\tilde{\mathbf{F}}$ is generated by concatenating all the $\tilde{\mathbf{F_{c}}}$ together:
\begin{equation}
\tilde{\mathbf{F_{c}}}=\frac{\sum_{t=1}^{L}P(c|\mathbf{y_{t}},\lambda)(\mathbf{y_{t}}-\mathbf{\mu_{c}})}{\sum_{t=1}^{L}P(c|\mathbf{y_{t}},\lambda)}.
\label{eq_3}
\end{equation}
Then the centered mean supervector $\tilde{\mathbf{F}}$ is projected as follows:
\begin{equation}
\tilde{\mathbf{F}}\to \mathbf{Tx},
\label{eq_4}
\end{equation}
where $\mathbf{T}$ is a rectangular low rank total variability matrix and $\mathbf{x}$ is the so-called i-vector \cite{dehak2010front}.

\subsubsection{The MFCC i-vector}

The MFCC i-vector is extracted by the aforementioned i-vector framework with the acoustic level MFCC features.
For cepstral feature extraction, a 25ms Hamming window with 10ms shifts was adopted. Each utterance was converted into a sequence of 36-dimensional feature vectors, each consisting of 18 MFCC coefficients and their first order derivatives. We employed the English phoneme recognizer \cite{schwarz2006hierarchical} to perform the voice activity detection (VAD) by simply dropping all frames that are decoded as silence or speaker noises.

\subsubsection{The MFCC-PPP i-vector}

It is reported in \cite{li2014interspeech,DHaro2014} that by combining the phonetic level phoneme posterior probability based tandem features with the acoustic level MFCC features at the feature level, the performances on speaker verification and  language identification are significantly enhanced. In this work, the MFCC-PPP i-vector is extracted the same way as  in \cite{li2014interspeech} following  the generalized i-vector framework.
We employed the multilayer perceptron (MLP) based phoneme recognizer \cite{schwarz2006hierarchical} with a provided English acoustic model trained on the TIMIT database to perform the phoneme decoding. The GMM model size and the tandem feature dimensionality are 512 and 32, respectively. 

\subsubsection{The MGDCC-PPP i-vector}

\begin{table*}[t]
    \footnotesize
    \centering
    \begin{tabular}{|c|c|c|c|c|c|c|c|}
        \hline
        System & \diagbox{\ \ \ \ \ \ \ \ \ \ \ \ \ \ \ \ \ \ \ \ \ \  Feature}{EER}  & LIBLINEAR & LIBPOLY & COSINE SCORING & KNN & Simplified & two stage \\
               & \diagbox[dir=SE]{classification method}{}&&&&& PLDA& PLDA\\\hline
             1 & MFCC i-vector & 8.46 & 6.63& 16.1 & 9.95 & 12.01 & 17.84\\\hline
             2 & PPP i-vector & 1.72& 1.26& 3.6& 3.4 & 2.29 & \\\hline
             3 & MFCC-PPP i-vector & 1.86 & \textbf{1.06} & 2.86 & 2.46 & 1.89 & 10.18  \\\hline
             4 & MGDCC-MFCC-PPP i-vector & 2.97 & 2.06 & 6.52& 3.43 & 3.95 & 17.79 \\\hline
            5  &            OPENSmile & 2.03& 1.57 &  & & & \\\hline
             6 & Fusion 1+2+3+4 & &  & 1.63 & 1.37 & 1.09 &\\\hline
             7 & Fusion 1+2+3+4+5 & 0.54& \textbf{0.29} &&& & \\\hline
    \end{tabular}
        \caption{Performance of the proposed methods on the development data} \label{dev-res}
\end{table*}

\begin{table*}[t]
    \footnotesize
    \centering
    \begin{tabular}{|c|c|c|c|c|c|c|c|c|c|c|c|}
        \hline
        & S1 & S2 & S3 & S4 & S5 & S6 & S7 & S8 & S9 & S10 & Average\\\hline
     Fusion 1+2+3+4+5-LIBPOLY   &  0.1137& 1.0332& 0.0482& 0.0412& 0.6614& 0.7112& 0.2297& 0.0108& 0.1336& 29.6649& 3.265 \\\hline
    \end{tabular}
        \caption{Performance of the fusion systems with different spoofing conditions on the testing data} \label{test-res}
\end{table*}

The MGDCC-PPP i-vector is calculated the same way as the MFCC-PPP i-vector except that here we replace the acoustic level MFCC features with the phase domain MGDCC features. The MGDCC feature is a kind of frame-level feature focusing on the speech phase characteristics. It has been shown that phase domain features are effective for anti-spoofing countermeasures \cite{DBLP:conf/icassp/WuXCL13}. In order to calculate the MGDCC feature, we need to obtain the modified group delay function phase spectrum (MGDFPS) \cite{DBLP:conf/icassp/ZhuP04} first.

Given the data $x_n$ of a short time window, the MGDFPS spectrum $\tau_{\rho,\gamma}(\omega)$ is calculated as follows \cite{DBLP:conf/icassp/ZhuP04}:
\begin{equation}
\tau_{\rho}(\omega) = \frac{X_{R}(\omega)Y_{R}(\omega) + Y_{I}(\omega)X_{I}(\omega)}{|S(\omega)^{2\rho}|} 
\end{equation}
\begin{equation}
\tau_{\rho,\gamma}(\omega)=\frac{\tau_{\rho}(\omega)}{|\tau_{\rho}(\omega)|}|\tau_{\rho}(\omega)|^{\gamma}
\end{equation}
where $X(\omega)$ and $Y_(\omega)$ are the fourier transforms of speech signal $x(n)$ and $nx(n)$; $X_{R}(\omega)$ and $X_{I}(\omega)$ are the real and imaginary parts of $X(\omega)$; $Y_{R}(\omega)$ and $Y_{I}(\omega)$ are the real and imaginary parts of $Y(\omega)$, respectively. $|S(\omega)|^2$ is calculated by applying a smoothing over $X(\omega)$ \cite{DBLP:conf/icassp/ZhuP04}. After applying the Mel-frequency filter banks and Discrete Cosine Transform, MGDCC feature is obtained. More details can be found in \cite{DBLP:conf/icassp/WuXCL13}.

\subsubsection{The OpenSMILE feature vector}
The OpenSMILE feature  is a 6373 dimensional utterance level feature vector extracted by the OpenSMILE toolkit \cite{eyben2010opensmile} using the configuration file provided by the 2014  Paralinguistic Challenge \cite{Paralinguisticsis2014}. Since various kinds of features, such as MFCC, loudness, auditory spectrum, voicing probability, F0, F0 envelop, jitter, and shimmer, etc., are included, this feature set can capture spoofing information at both the acoustic and prosodic levels. In our system, it served as a baseline as well as a supplement to those i-vector subsystems.

\subsection{Back-end modeling}\label{sec:modeling}
After feature vectors are extracted, we apply different classification methods for the back-end modeling.
\subsubsection{The K-nearest neighbor classification (KNN)}

KNN is a non-parametric multi-class classifier. The utterances in the training set are divided into human set and spoofed set. For each test utterance $x_t$, $K$ nearest neighboring utterances are found in the training set and the score is calculated based on the class distribution of these K nearest neighbors.

\subsubsection{The cosine similarity scoring}

The cosine similarity between two vectors is calculated as follows:
\begin{equation}
similarity(\mathbf{x}, \mathbf{y}) = \frac{\mathbf{x}^{t}\mathbf{y}}{||\mathbf{x}||_{2} ||\mathbf{y}||_{2}}
\end{equation}

In our system, a mean vector of all the human utterances in the training data set is calculated. For each test utterance, the score is computed as the cosine similarity between itself and the human class mean vector.

\subsubsection{PLDA modeling}

We first applied the simplified PLDA modeling \cite{DBLP:conf/interspeech/Garcia-RomeroE11} as the back-end assuming that there are six special speakers (five spoofing channels plus one human channel), each represents a spoofing type or the original genuine speech. Furthermore, we also adopted the two subspace (speaker subspace and spoofing subspace) PLDA presented in \cite{prince2007probabilistic} to model the i-vectors. The standard log likelihood ratio based hypothesis is emploied for the scoring \cite{DBLP:conf/interspeech/Garcia-RomeroE11,prince2007probabilistic}.

\subsubsection{Support Vector Machine}
We formed the anti-spoofing countermeasure as a two class classification task for  SVM modeling.
The linear kernel LIBLINEAR \cite{Fan:2008:LLL:1390681.1442794}  and its polynomial kernel extension LIBPOLY \cite{libpoly} are adopted as the back-end SVM classifiers and we applied the min/max normalization (range -1 to +1) for each feature dimension on the training, development and test sets with parameters computed only from the training data.

\subsubsection{Score fusion}

We simply employed the weighted summation fusion approach at the score level to further enhance the performance. The fusion  weights were tuned on the development data set.

\section{Experimental results} \label{sec-ex}

The results of our four subsystems on the development data are shown in the Table \ref{feature-fuse}. We can observe that feature level fusion with PPP feature improves the performance. Compared to the MFCC i-vector subsystem (EER = 6.63\%), the EER of MFCC-PPP i-vector subsystem is reduced to 1.06\%. On the other hand, the openSMILE feature outperformed the MFCC i-vector subsystem which might be due to the inclusion of prosodic level information.

Furthermore, to obtain a robust countermeasure system, different backend classification techniques were evaluated. Table \ref{dev-res} shows the performance on the development data. Among these six classification methods, LIBPOLY achieves the best performance with 0.29\% EER on the development data. The improvement of LIBPOLY against LIBLINEAR motivated us to further increase the SVM polynomial kernel degree. Table \ref{svm-res} shows that SVM with high degree polynomial kernel may lead to overfitting. 

With regard to PLDA backends, it shows that the simplified PLDA tends to be more robust against those unseen spoofing attacks. As shown in Table \ref{plda-svm}, we simulated unknown spoofing attacks by using four kinds of spoofed utterances in the training and the remaining one in the testing. Although its performance was as good as LIBLINEAR against familiar spoofing attacks, it outperformed LIBLINEAR on the unseen testing data, especially where the unknown attacks were related to speech synthesis (index 3 and 4). The two stage PLDA only achieved moderate results in Table \ref{dev-res} which might be because total speakers number in the training data is limited (25) and the speaker subspace may not be orthogonal to the spoofing subspace. 

Table \ref{test-res} presents our fusion system results with each individual spoofing condition on the test data. Here S1 to S5 are know attacks and S6 to S10 are unknown attacks. Our system performed well on all attacks except S10, on which most challenge participants got unsatisfied results. 

Finally, our fusion system (system 7) achieved 0.38\% and 6.15\% EER against known and unknown attacks, respectively.

\begin{table}[t]
    \centering
    \begin{tabular}{|c|c|}
        \hline
        Methods & EER(\%) \\\hline
        MFCC i-vector & 6.63 \\\hline
   MFCC-PPP i-vector & \textbf{1.06} \\\hline
        MGDCC-PPP i-vector & 2.23 \\\hline
        OpenSmile & 1.57 \\\hline
    \end{tabular}
    \caption{Performance of the four subsystems on the development data} \label{feature-fuse}
\end{table}

\begin{table}[t]
    \footnotesize
    \centering
    \begin{tabular}{|c|c|c|c|c|c|}
        \hline
       polynomial & 1 & 2 & 3 & 4 & 10\\
                         kernel degree& (LIBLINEAR)&(LIBPOLY) & & & \\
                              \hline
           EER & 1.86 & 1.06 & 1.03 & 1.00 & 2.32 \\\hline
    \end{tabular}
    \caption{Performance of the MFCC-PPP i-vector SVM subsystems with different polynomial kernel degrees} \label{svm-res}
\end{table}

\begin{table}[t]
    \footnotesize
    \centering
    \begin{tabular}{|c|c|c|c|}
        \hline
        train set & test set & PLDA &LIBLINEAR \\\hline
        human+spoof[2,3,4,5] & human+spoof[1] & 3.57 & 3.4  \\\hline
        human+spoof[1,3,4,5] & human+spoof[2] & 4.8 & 7.69 \\\hline
        human+spoof[1,2,4,5] & human+spoof[3] & 0.2 & 0.71 \\\hline
        human+spoof[1,2,3,5] & human+spoof[4] & 0.2 & 0.66  \\\hline
        human+spoof[1,2,3,4] & human+spoof[5] & 4.49 & 11.81 \\\hline
    \end{tabular}
    \caption{Performance of the LIBLINEAR and the simplified PLDA backends on the unknown spoofing testing conditions} \label{plda-svm}
\end{table}

\section{Conclusions} \label{sec-con}

This paper presents an anti-spoofing countermeasure system based on a multi-feature and multi-subsystem fusion approach. By fusing the phonetic level phoneme posterior probability tandem features with  the acoustic level MFCC features or the phase level MGDCC features, the system performance is significantly enhanced. Combining the proposed i-vector subsystems with the OpenSMILE baseline which covers the acoustic and prosodic level information further improves the final performance. For the back-end modeling, two classes support vector machine outperforms the one class cosine similarity or PLDA scoring on the development data where the spoofing attack types are known. The one class scoring method achieves more robust performance on the unseen testing data where the spoofing conditions are unknown.

  \ninept
  \bibliographystyle{IEEEtran}
  \bibliography{SYSU}

\end{document}